\documentclass[amsmath,amssymb,floatfix,superscriptaddress]{iopart}
\usepackage{graphicx}

\begin{document}

\newcommand{\tfrac}[2]{\ensuremath{\textstyle \frac{#1}{#2}}}

\title{Fragmented superfluid due to frustration of cold atoms in
  optical lattices}

\date{\today}

\author{Juan Jos\'e Garc{\'\i}a-Ripoll }
\address{Max-Planck-Institut f\"ur Quantenoptik,
  Hans-Kopfermann-Str. 1, Garching b. M\"unchen, D-85748, Germany}

\address{Universidad Complutense, Facultad de CC. F\'{\i}siscas, Ciudad
  Universitaria s/n, Madrid E-28040, Spain}
\ead{juanjose.garciaripoll@gmail.com}

\author{Jiannis K. Pachos}
\address{Department of Applied Mathematics and Theoretical Physics,
University of Cambridge, Wilberforce Road, Cambridge CB3 0WA, UK}

\address{Quantum Information Group, School of Physics \& Astronomy,
University of Leeds, Leeds LS2 9JT UK}

\begin{abstract}
  A one dimensional optical lattice is considered where a second dimension
  is encoded in the internal states of the atoms giving effective ladder
  systems.  Frustration is introduced by an additional optical lattice that
  induces tunneling of superposed atomic states. The effects of
  frustration range from the stabilization of the Mott insulator phase with
  ferromagnetic order, to the breakdown of superfluidity and the formation
  of a macroscopically fragmented phase.
\end{abstract}

\submitto{\NJP}

\maketitle

\section{Introduction}

One of the most interesting open fields in quantum magnetism is the study of
frustration \cite{diep04}. Frustrated models are described by Hamiltonians
with competing local interactions such that the ground state cannot minimize
their energy simultaneously. Frustration can appear due to the geometry of a
problem, the Ising model on a triangular lattice being a paradigmatic case,
or due to the coexistence of ferro- and antiferromagnetic interactions, as
in spin glasses. Frustrated models typically have highly degenerate ground
states, which can become ordered by increasing the temperature or by quantum
fluctuations --- i.~e. ``order by disorder''. Theoretical and numerical
problems, such as the large dimensionality or the sign problem in Monte
Carlo simulations make it very difficult to study frustrated Hamiltonians.

Quantum simulation has been put forward as a tool to probe the physics of a
variety of many-body systems. In particular, schemes with cold atoms in
optical lattices have been suggested to simulate arbitrary spin models
\cite{jane03,kuklov03,duan03,garcia-ripoll03b,yip03,garciaripoll04} and
some interesting frustrated Hamiltonians \cite{santos04,polini05}. Compared
to the alternative of looking for or inducing low-dimensional behavior on
existing magnetic materials \cite{gopalan94,azuma94} and in organic
conductors \cite{miller94,macedo95}, the cold atoms and molecules offer
greater flexibility in terms of variable geometry and interaction
strength. However, the majority of current proposals are perturbative and
their effective interactions are rather weak. This makes their experimental
realization challenging, as it requires very low temperatures. A solution
suggested to solve the problem of weak interactions is to replace the atoms
with polar molecules \cite{micheli05}.

In this paper we follow a different approach. First of all, spin up and
spin down states are identified with particle and hole configurations. This
gives naturally an XX coupling which is proportional to the tunneling
amplitude \cite{paredes04} and, thus, it is strong and less experimentally
demanding. By using the internal state of the atoms to encode a second
virtual dimension and applying a spatially dependent Raman coupling, we are
able to induce in-plane frustration. The result is a new frustrated model,
which contrasts with related literature on spin ladders, where the
Hamiltonian either use antiferromagnetic isotropic
\cite{hung06,vekua06,allen00,kim00,nerseyan98,gelfand91} or XXZ
\cite{haldane82} interactions, or use purely XY interactions but require a
different lattice geometry \cite{hakobyan01}, or both. More important, the
problem we consider here is a full Bose-Hubbard Hamiltonian, and the main
result is that the transport properties of the lattice change dramatically,
causing the breakdown of superfluidity even for large densities in which
the analogy to spin models no longer applies. In particular, in the most
interesting case the Mott insulator is replaced by a new dimerized phase
made, where neighboring pairs of sites have maximal coherence and there is
a fast decay of coherence between pairs of sites as a function of
distance. This new phase is reminiscent of a Bose glass~\cite{fisher89},
with the important difference that it has been induced by frustration and
not by disorder. Our statements are supported by a variety of analytical,
variational and numerical solutions.

The paper is organized as follows. In Sect.~\ref{sec:model} we introduce
the two Hubbard models that we are going to use, explain how they are
implemented with cold atoms in optical lattices and briefly justify the use
of the word ``frustration''. In Sect.~\ref{sec:diagonal} we focus on a
model with diagonal interactions across each square plaquette. We first
discuss the different phases as obtained from DMRG calculations, including
qualitative arguments about why these phases are expected. Afterwards we
explain in more detail how fragmentation happens and what happens in the
limit of strong interactions, and what are the correlation properties of
all available phases. In Sect.~\ref{sec:villain} we briefly present a model
with frustrated rectangular lattice. In Sect.~\ref{sec:experiments} we
study in great detail all issues regarding the implementation of our ideas
in current experiments with optical lattices. First we do a microscopic
derivation of the parameters in the Hubbard Hamiltonians, including all
possible sources of error. The main conclusion is that the requirements for
studying these frustrated models appear to be within reach of current
experiments and that there are no side-effects from introducing a Raman
coupling in the experiment. We explain how the different phases studied in
this paper could be detected, either from time of flight measurements or
more sophisticated correlation measurements. Finally, we discuss possible
sources of imperfection, such as temperature or a residual harmonic
confinement. The last section (Sect.~\ref{sec:conclusions}) contains a
brief summary of our main results and possible implications and connections
to other works.

\section{The model}
\label{sec:model}

\subsection{Bose-Hubbard model}

The system that we are studying is that of cold atoms confined by an
off-resonance optical lattice that forms a 1D bosonic lattice gas
\cite{paredes04}. As in recent experiments \cite{widera04}, the
lattice traps atoms in two internal states.  If the confinement is
strong, the effective model will be a Hubbard Hamiltonian
\cite{jaksch98}
\begin{eqnarray}
  H = -J\sum_{\langle i,j\rangle,\sigma} a_{i\sigma}^\dagger
  a_{j\sigma} + \sum_{i\sigma\sigma'}
  \frac{U_{\sigma\sigma'}}{2} n_{i\sigma}n_{i\sigma'}
  + H_R,
  \label{Hamiltonian}
\end{eqnarray}
where $J$ is the hopping amplitude and we assume that the on-site atomic
interactions are both repulsive and symmetric, $U_{\sigma\sigma'}= U > 0$.
Note that in this work, the internal state of the atoms are denoted by
Greek letters, $\sigma=\uparrow, \downarrow$; the site indices are denoted
by Roman characters, $i,j$, and go from $1$ to $L$ which is the length of
the lattice; finally $N:=N_\uparrow + N_\downarrow$, represents the total
number of particles.

\begin{figure}
  \centering%
  \includegraphics[width=0.8\linewidth]{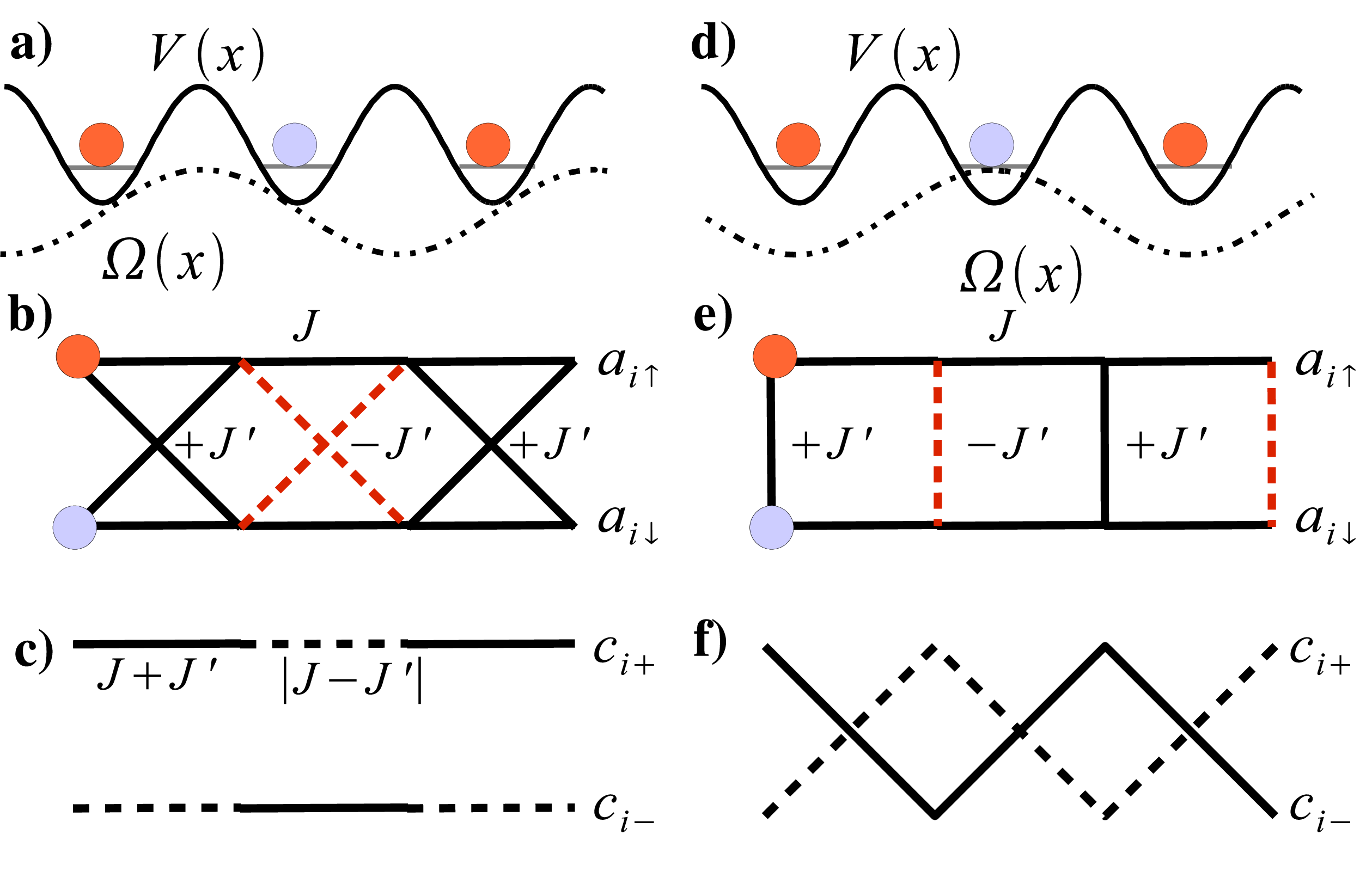}%
  \caption{(a) Raman-assisted tunneling. A second optical lattice
    (dash-dot line) connects neighboring Wannier functions of atoms in
    \textit{different} states with amplitude $J'$. On the other hand,
    normal hopping moves atoms while preserving the state, $J$. (b)
    The hard-core bosons limit resembles a spin ladder with
    frustrating diagonal interactions. (c) The change of variables in
    Eq.~(\ref{rotation}) replaces the frustrating terms with weak and
    strong tunnelling. (d-f) Similar drawings for the Villain model in
    Eq.~(\ref{villain}).}
\label{fig:lattice}
\end{figure}

The new ingredient in our Hamiltonian are two counter-propagating laser beams
that induce Raman transitions between the two atomic states. By adjusting
the phase, the polarization and the alignment of these beams, we can ensure
that the effective Rabi frequency, $\Omega(x)$, forms a lattice with twice
the period of the confinement [Fig.~\ref{fig:lattice}].  We can interpret
our one-dimensional system as having two 1D lattices, one for each internal
state of the atom. Both chains are coupled forming a ladder thanks to both
the interaction, $U$, and to $H_R$, a Raman term with the spatially
dependent Rabi frequency, $\Omega(x)$.

We shall consider two specific configurations of the Raman coupling.
When its maxima and minima coincide with those of the confining
lattice [Fig.~\ref{fig:lattice}d], the result will be on-site Rabi oscillations
\begin{equation}
  \label{villain}
  H_R = \sum_i (-1)^i J' (a^{\dagger}_{i\uparrow} a_{i\downarrow}
  + \mathrm{H.c.}).
\end{equation}
Alternatively, when they are displaced by half a period we will have
diagonal interactions [Fig.~\ref{fig:lattice}a]
\begin{equation}
  H_R = \sum_i (-1)^i J' (a^{\dagger}_{i+1\uparrow} a_{i\downarrow} +
  a^{\dagger}_{i+1\downarrow} a_{i\uparrow}+ \mathrm{H.c.}).
  \label{diagonal}
\end{equation}

\subsection{Origin of frustration}
\label{sec:frustration}

The frustration induced by $H_R$ becomes evident in the limit of
hard-core bosons, $a^2_{i\sigma}=0$ \cite{paredes04}.  Identifying the
bosonic operators, $a_{i\uparrow}$ and $a_{i\downarrow}$, with the
spin operators $\sigma^{-}_{i1}$ and $\sigma^{-}_{i2}$, the Hamiltonian
(\ref{Hamiltonian}) becomes an XX model on a spin ladder
[Fig.~\ref{fig:lattice}b,e]. Thus, while the hopping $J$ translates into
a ferromagnetic XX interaction along the legs, the Raman coupling
$H_R$ is a transverse XX interaction which alternates from ferro- to
antiferromagnetic.

Take for instance the diagonal interactions (\ref{diagonal}). The
associated spin model is
\begin{eqnarray}
  H &=& \sum_{i=1,2}\sum_{j=1\ldots N}
  J (\sigma_{ij}^{x} \sigma_{ij+1}^{x} + \sigma_{ij}^{y} \sigma_{ij+1}^{y}) +
  \nonumber\\
  &+& \sum_{j=1\ldots N} (-1)^jJ' (\sigma_{1j}^{x} \sigma_{2j+1}^{x} +
  \sigma_{1j}^{y} \sigma_{2j+1}^{y}) + \nonumber\\
  &+& \sum_{j=1\ldots N} (-1)^jJ' (\sigma_{2j}^{x} \sigma_{1j+1}^{x} +
  \sigma_{2j}^{y} \sigma_{1j+1}^{y}).
  \label{spin-model}
\end{eqnarray}
The frustration arises from the competition of positive and negative
contribution terms. For instance, looking at the sites $(1,j), (1,j+1),
(2,j+1)$ and $(1,j+2)$: the bonds on this triangle have three ferro- and
one antiferromagnetic terms, and there is no way in which the spins can be
aligned so as to minimize the energy of all terms [see also
Fig.~\ref{fig:lattice}b].

While the previous analogy is somewhat pleasing, one may wonder
whether having more than one particle per site or working with a
system which is not properly 2D will give rise to a trivial Physics.
Regarding the first point, when we have more than one atom per site
our model becomes equivalent to an array of Josephson junctions
\cite{polini05} where the frustration is still present and lays
on the choice of phase for
each individual site. Regarding the second point, some of the best
studied models in frustrated quantum magnetism, such as the
Majumdar-Ghosh \cite{majumdar69,majumdar69b} and the zig-zag chain
\cite{haldane82} are also quasi-1D models. As we will show in the following
sections, the competition between the different terms in
Eq.~(\ref{Hamiltonian}) does indeed produce new and interesting effects with
the advantage of being analytically tractable.

\section{Diagonal interactions}
\label{sec:diagonal}

In contrast to fully two-dimensional systems, in our quasi-1D models can be
greatly simplified by rewriting the frustrated coupling using dressed states
\begin{equation}
c_{i\pm} :=
\frac{1}{\sqrt{2}}(a_{i\uparrow} \pm a_{i\downarrow}). \label{rotation}
\end{equation}
We will begin by studying the Raman interaction~(\ref{diagonal}). With the
previous change of variables, our Hamiltonian becomes
\begin{equation}
  H = -\sum_{i;\sigma=\pm}
  [J+\sigma(-1)^{i}J'] (c^\dagger_{i\sigma} c_{i+1\sigma} + \mathrm{H.c.})
  + \frac{U}{2} n_{i}^2,
  \label{modelc}
\end{equation}
where $n_{i}:=c_{i+}^\dagger c_{i+} + c_{i-}^\dagger c_{i-}$ is the total
population of a single site. Eq.~(\ref{modelc}) models a ladder whose legs
are made of junctions with strong, $J_{+}=J+J'$, and weak hoppings,
$J_{-}=|J-J'|$, as shown in Fig.~\ref{fig:lattice}c.  Notice that the model
is symmetric under the exchange of $J$ and $J'$. In the following we will
often exchange descriptions between $J_{+},J_{-}$ and $J,J'$, and we will
assume, without loss of generality, that all these parameters are not
negative.

\subsection{Phase diagram}
\label{sec:phases}

In this section we present zero temperature results for
Hamiltonian~(\ref{modelc}). We have computed the ground states and first
excitations of this model using the Density Matrix Renormalization Group
(DMRG) method in the Matrix Product States (MPS) formalism
\cite{verstraete04a,dmrg}. We have conducted accurate simulations for
$L=8,16,32, 50$ and $70$ sites, using a cutoff of 4 particles per
spin. These values give us a local dimension of the Hilbert space of 25
states, which is very big and can only be handled with state-of-the art
optimizations, such as working in sectors with well defined number of
particles or angular momenta. Due to the large size of the local Hilbert
space, we have typically worked with MPS of size $100$ and checked
convergence for different points with up to $D=200$\footnote{Note that due
  to our way of computing the two lowest excited states, which uses
  different MPS for the ground state and excitations. These simulations are
  comparable to DMRG calculations with $D=300$ and $D=600$ states,
  respectively, because, in the DMRG parlance, the effective basis states
  are optimized independently for each excited state.}. In our simulations
we have computed the ground state and lowest energy states for
$N/L=1/2,1,5/4,3/2,7/4,2$ particles, as well as the energy to add a
particle, $\mu_p := E_{N+1}-E_N$, or to create a hole,
$\mu_h:=E_{N}-E_{N-1}$, and verified that these values did not change with
a larger cutoff.

These zero temperature simulations reveal a very simple picture that we
summarize here. As illustrated in Fig.~\ref{fig:lattice}d, when we increase
$U$ for fixed $J_+$ and $J_-$, the system first experiences a phase
transition from a superfluid to a fragmented or dimerized phase in which
coherence is maximal between pairs of neighboring sites. As the
interaction is further increased these fragments evolve smoothly (i.~e. a
crossover) to a Mott insulator.  While the first phase is gapless and its
excitations are phonons, the crossover contains a gapped phase whose lowest
excitations are localized spin flips.

\begin{figure}[t]
  \includegraphics[height=6.0cm]{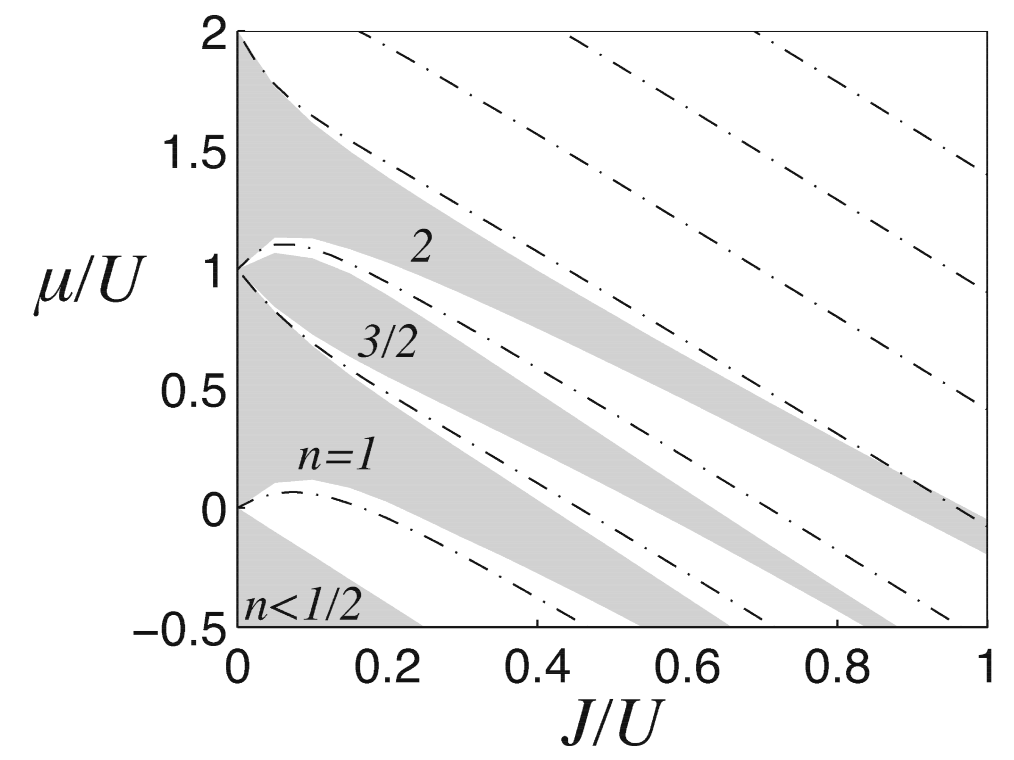}
  \includegraphics[height=5.7cm]{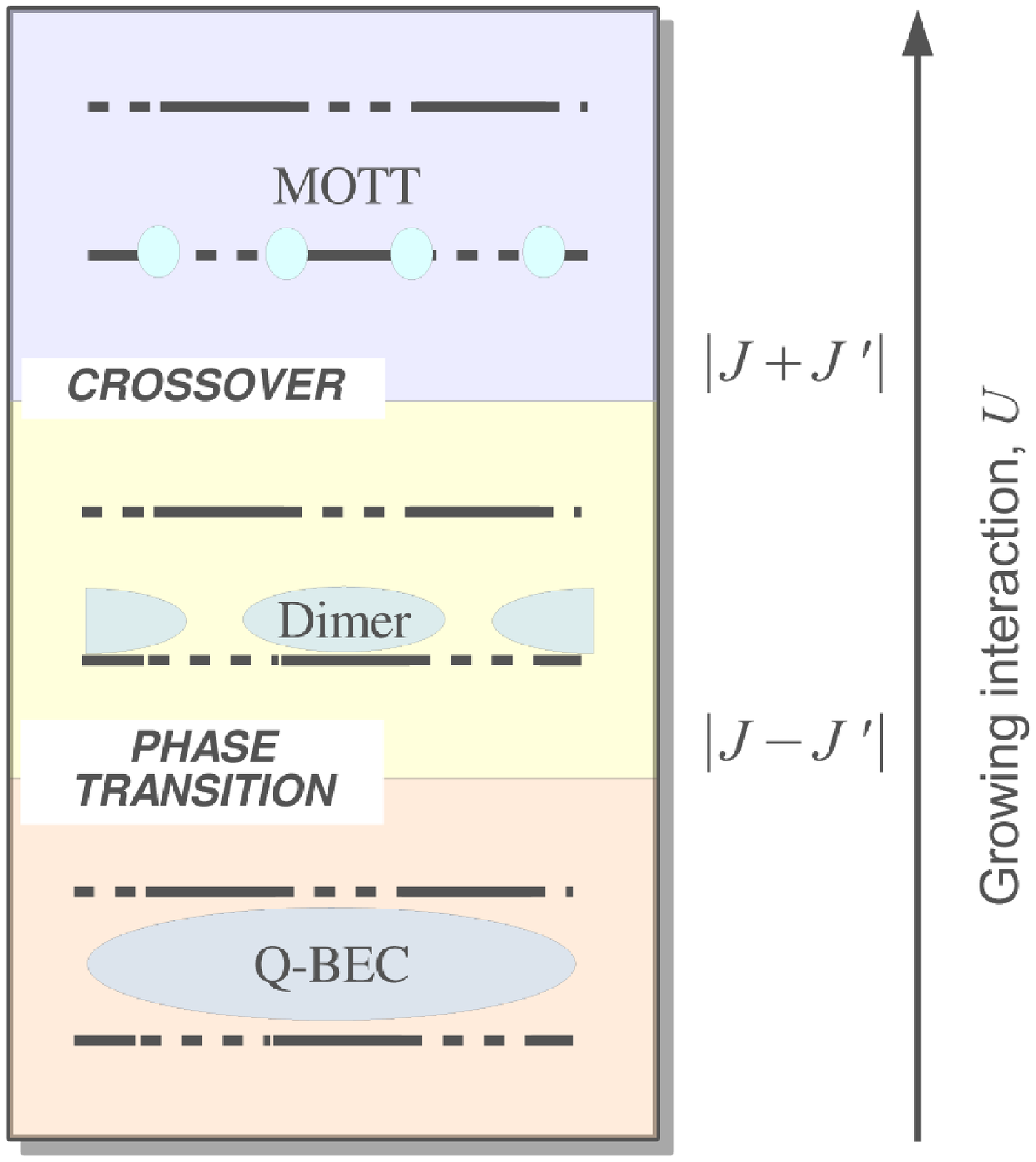}%
  \caption{(left) Phase diagram for the diagonal interactions
    (\ref{diagonal}) computed using DMRG. For $J'=0.9J$, the gray areas
    denote the incompressible Mott lobes in the space of chemical potential
    ($\mu/U$) vs. hopping amplitude ($J/U$). The in between regions are in
    superfluid phase. For $J'=J$, the phase space fills with incompressible
    regions, each one having integer or half-integer filling,
    $N/L=1,3/2,2,...$. These regions are delimited by the dash-dot
    lines. (right) Schema of the different phases as a function of the
    interaction $U$.}
  \label{fig:dmrg}
\end{figure}

In the following sections we will study in detail the properties of the
different coupling regimes, numerically as well as analytically where
possible. Nevertheless it is possible to obtain a qualitatively picture of
the three regimes and the associated ground states by variational
wavefunctions. First of all, for weak interactions, $U\ll J_{\pm}$, the
ground state is a uniform superfluid that spread over the lattice
\begin{equation}
  \label{bec}
  |\psi_{\mathrm{SF}}\rangle \propto \frac{1}{\sqrt{N!}}
  \left [\sum_i ( \alpha a_{i\uparrow}^\dagger
    + \beta a_{i\downarrow}^\dagger)\right]^N |\mathrm{vac}\rangle,
\end{equation}
with the usual rotational symmetry on the $(\alpha,\beta)$ space.

For stronger interactions, $J_{-} < U/4 \ll J_{+}$, the energy of
$|\psi_{\mathrm{SF}}\rangle$ is larger than a state made of fragmented
condensates which reside on the junctions with large hopping, $J_{+}$
[Fig.~\ref{fig:lattice}c],
\begin{eqnarray}
  |\psi_{\mathrm{frag}}\rangle &\propto& \prod_{k}
  (c_{2k,+}^\dagger + c_{2k+1,+}^\dagger)^{n_+}\times
  (c_{2k+1,-}^\dagger + c_{2k+2,-}^\dagger)^{n_-}
  |\mathrm{vac}\rangle.
  \label{fragmented}
\end{eqnarray}
Here, for integer filling, $N/L=1,2,3,\ldots$, the ground state has is one
with spontaneously broken rotational symmetry, a ferromagnetic state with
either $n_+=0$ or $n_-=0$. This effect arises from the contributions to the
energy of both the interaction $U$ and the hopping $J_{-}$, as it will be
explained in Sect.~\ref{sec:fragmented}.

Another symmetry break happens when the interactions become dominant,
$U\gg J_{\pm}$.  Indeed, in the limit of strongly interacting bosons
and $N=L$, all ground states can be written in the form
\begin{equation}
  |\psi_{\mathrm{Mott}}\rangle \propto \prod_i (c_{i+}^\dagger)^{n_{i+}}
  (c_{i-}^\dagger)^{n_{i-}}|\mathrm{vac}\rangle.
  \label{localized}
\end{equation}
The quantum fluctuations generated by the hopping select, among all
disordered ground states, one particular ferromagnetic state with
either $n_{-}=0$ or $n_{+}=0$. This case is further discussed in
Sect.~\ref{sec:hard-core}.

In the space of parameters, $J/U$, vs. chemical potential, $\mu$, we can
draw a phase diagram similar to that of the Mott-insulator transition
\cite{fisher89}. As Fig.~\ref{fig:dmrg} shows, for $J'=0.9J$, the curves of
particle and hole excitations, $\mu_p$ and $\mu_h$, create lobes (gray
areas) containing fragmented phases. Crossing these lines amounts to a
quantum phase transition to a superfluid, on the exterior of the lobes.
Qualitatively, these lobes are similar to those of the Bose-Hubbard model
\cite{fisher89}, but about $J/|J-J'|$ times larger, so that for $J'=J$ the
lobes become infinitely large and all ground states are fragmented. In this
limit of $J'=J$ and for integer and half-integer filling, that is
$N/L=1,3/2,2,\ldots$, we have on phase space infinitely many incompressible
``stripes'' whose borders, $\mu_p$ and $\mu_h$, can be computed exactly
[See Sect.~\ref{sec:fragmented} and Eq.~(\ref{boundaries})]. As explained
above, within each stripe the state of the atoms changes smoothly from a
Mott insulator (small $J_{+}$) to fragmented or dimerized phase where atoms
delocalized between pairs of neighboring sites (large $J_{+}$).
Interestingly, in the macrocanonical ensemble, all phases with other
filling factors collapse to the lines between these regions. The large
degeneracies of these other fillings has made it impossible for us to study
systematically the nature of the those states and of their excitations.

\begin{figure}[t]
  \includegraphics[width=0.45\linewidth]{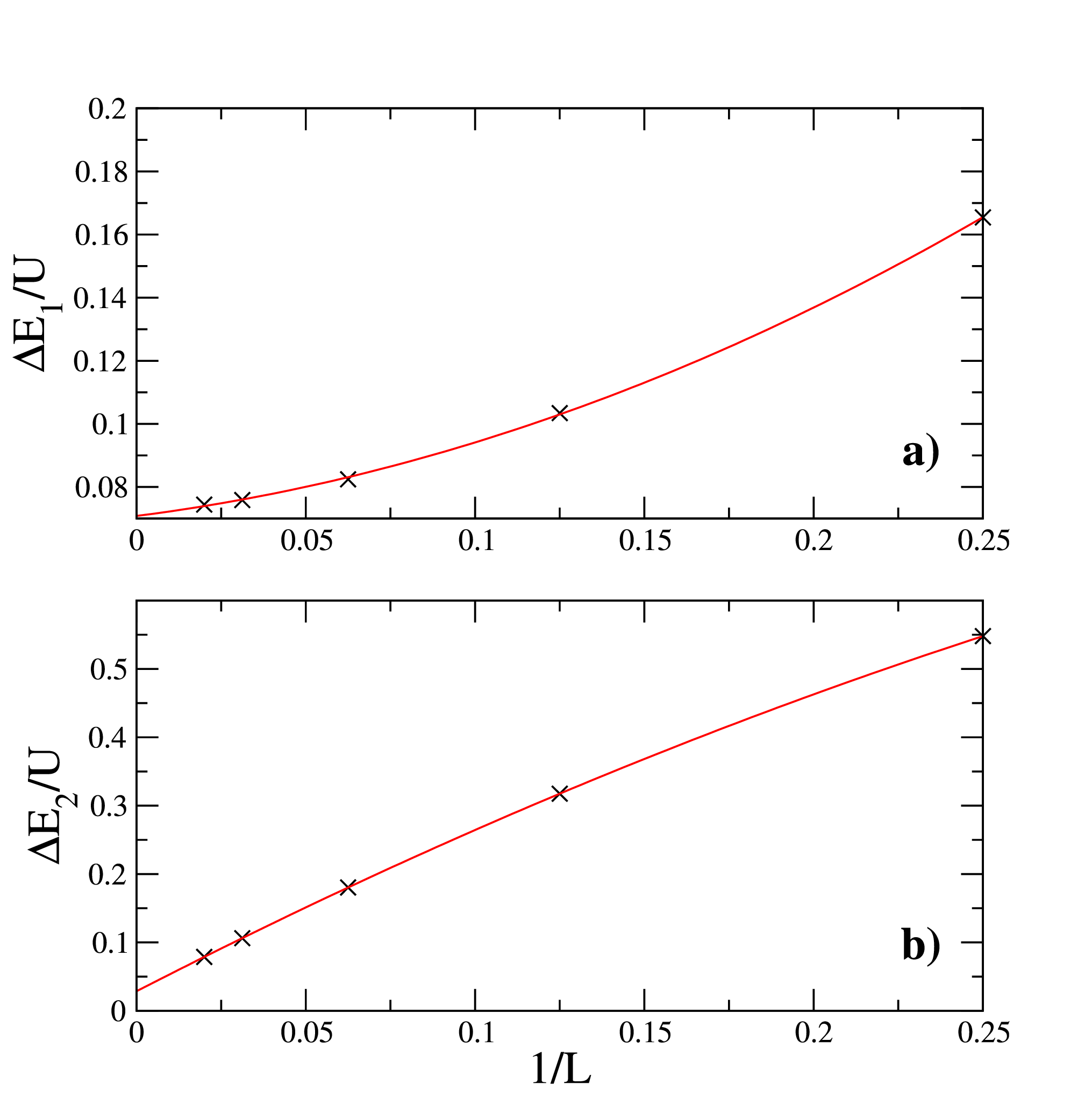}%
  \includegraphics[width=0.55\linewidth]{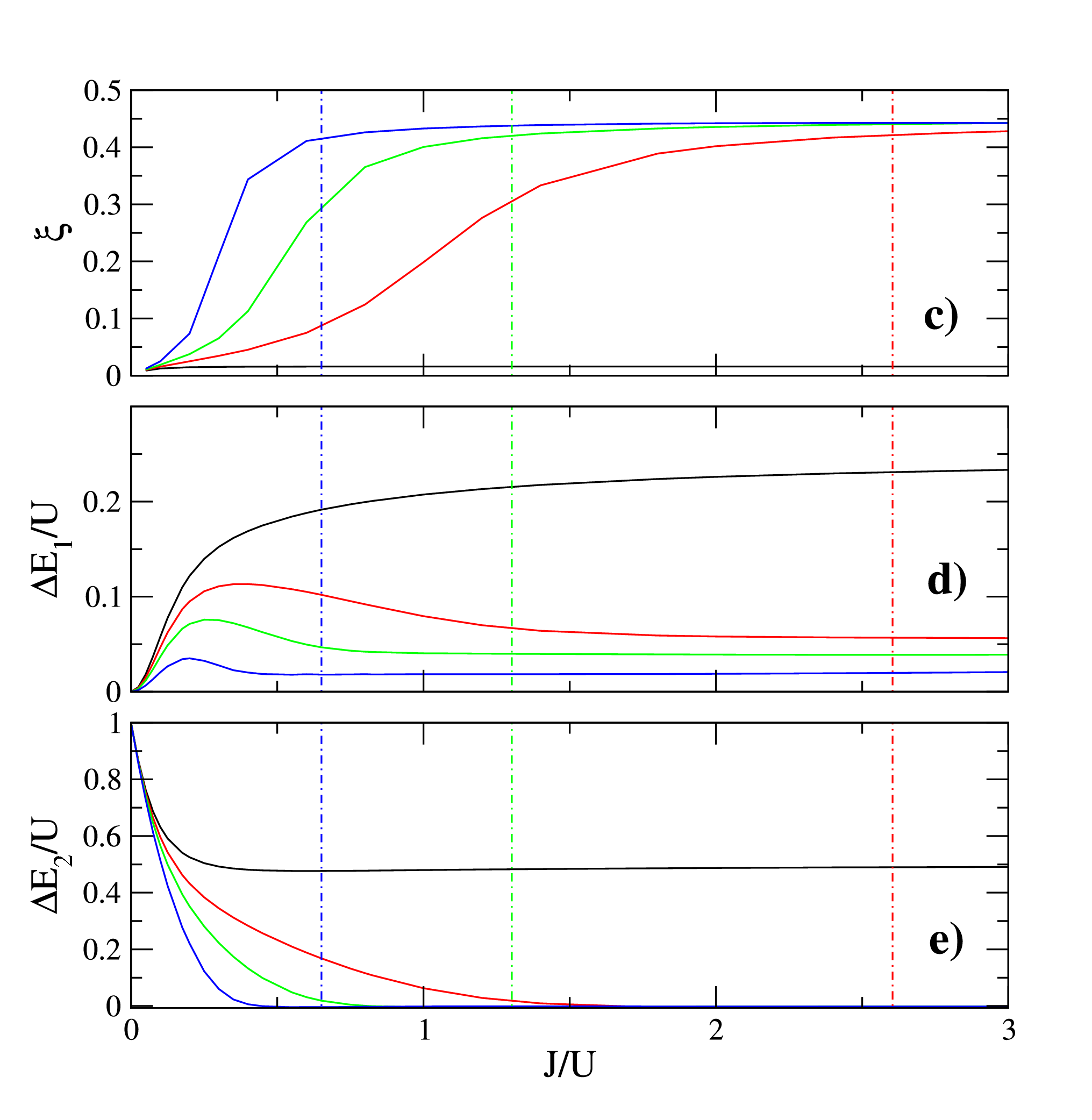}
  \caption{(a,b) Finite size scaling of the energy gaps for spin and
    density excitations, $\Delta E_1$ and $\Delta E_2$, for a particular
    case $J'/U=0.9$, $J/U=2.8$. We fit both curves to a polynomial in $1/L$
    for $L=4,8,16,32,50$ and $70$ and take the limit $1/L\to 0$. With this
    technique we plot in (d,e) the energy gaps vs. hopping amplitude, for
    $J'/J=1.0,0.9,0.8,0.6$ (top to bottom). With vertical lines we mark the
    points where the breakdown of superfluidity is roughly expected to take
    place, $U_{1D}\simeq 3.84 |J'-J|$. (c) Mean correlation length, $\xi
    \sim \sum_{i,\delta} |\delta| \langle a^\dagger_{i} a_{i+\delta}\rangle
    / \sum_{i,\delta}\sum \langle a^\dagger_{i} a_{i+\delta}\rangle$, for
    $J'/J=0.6,0.8,0.9,1.0$.}
  \label{fig:gap}
\end{figure}

To gain further insight on the properties of the ground state, let us focus
on the unit filling case. The transition from the Mott phase to the
superfluid is accompanied by the change of several excitation gaps.  One is
related to the energy required to add or remove a particle, the other one
is the energy gap required to create a density perturbation for fixed
number of particles. Both gaps close at the same point when we jump from
the fragmented phase to the uniform superfluid, signaling a phase
transition from an incompressible to a compressible phase and also the
establishment of algebraically decaying correlations
[Fig.~\ref{fig:gap}]. This transition occurs around the critical value of
the interaction for a single-component 1D Bose-Hubbard model
\cite{rapsch99,kuehner98,freericks96,scalettar91} with hopping $J_{-}$,
that is $U \sim U_{1D}:= 3.84 J_{-}$ (See Sect.~\ref{sec:gaps}). A third
energy gap is present for excitations that take a particle out of one of
the coherent fragments into a different spin state, while preserving the
total number of particles. This gap freezes the dynamics of the spin of the
atoms. As shown in Fig.~\ref{fig:gap}a-b, it is maximal in the fragmented
phase, with a maximum value of order $U/4$. For strong interactions it
grows as $J'^2/U$ and in the superfluid phase it should close. However
being $U\gg J_{pm}$ a region gapless with respect to density excitations,
the DMRG simulations are probably not accurate enough to reflect this fact.

\subsection{Fragmented state}
\label{sec:fragmented}

\begin{figure}[t]
  \includegraphics[width=0.6\linewidth]{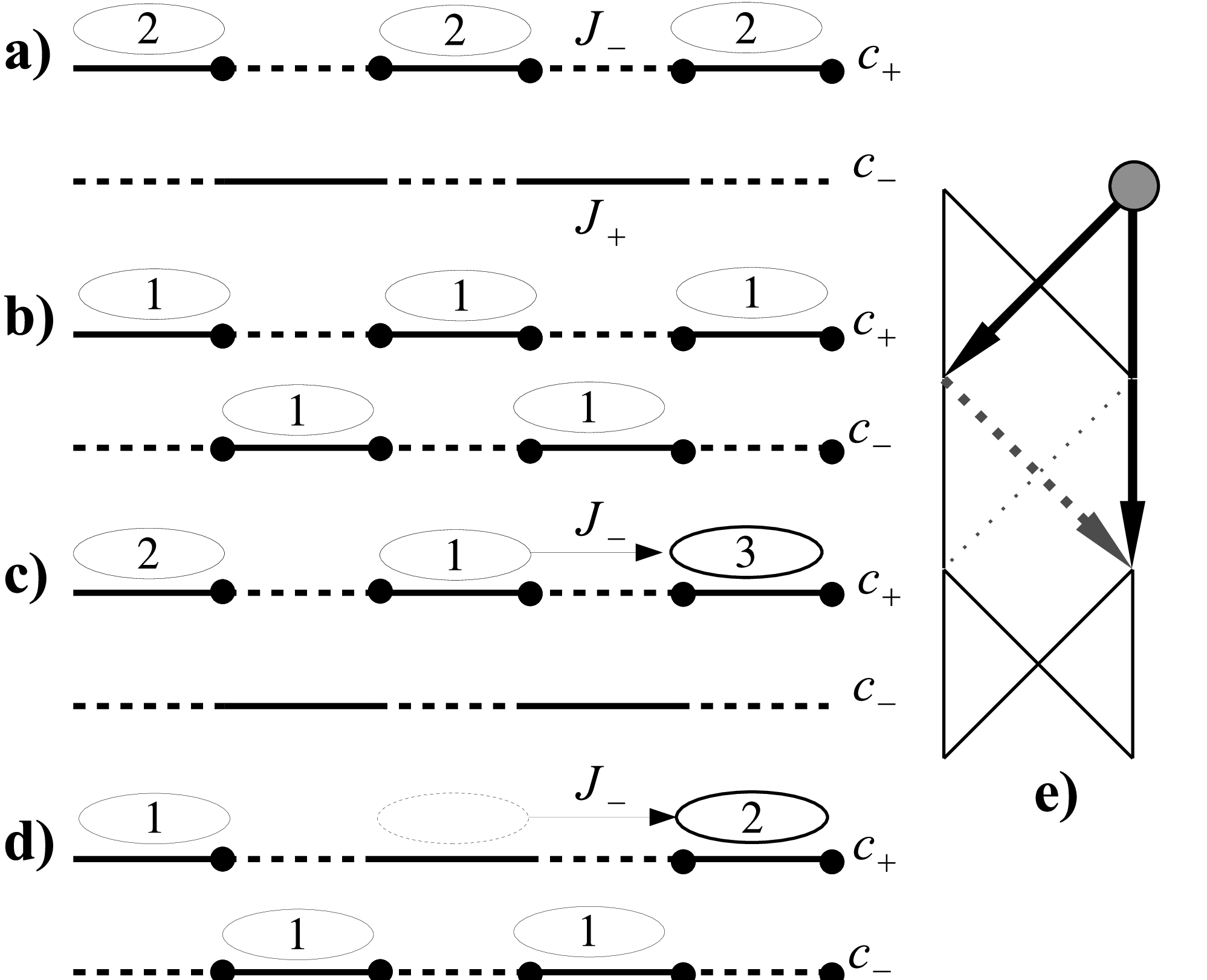}
  \caption{Two possible ground states for the fragmented phase:
    (a) with and (b) without spontaneous symmetry breaking. A nonzero
    value of $J_-$ couples the state (a) and (b) to the excitations
    (c) and (d), respectively. This coupling makes the state (a)
    preferred for any nonzero $J_-$ in the fragmented regime $J_- < U
    < J_+$. (e) A particle hopping through the lattice sees a kind of
    Mach-Zender interferometer. The antiferromagnetic hopping (dashed)
    gives rise to destructive interference on all paths longer than 2
    sites.}
  \label{fig:fragments}
\end{figure}

In this subsection we will comment on the configurations that appear for
$U\in [J_-,J_+)$. For simplicity we will begin with the case of balanced
hopping and Raman coupling and no interaction, $U=J_-=0$. Then it is
trivial to see that all ground states will be of the form
(\ref{fragmented}) with atoms delocalized on pairs of sites. In particular,
for unit filling $N=L$ we have three degenerate states: a uniform state
with all junctions populated $n_{\pm}=1$ [Fig.~\ref{fig:fragments}b] and
two states with broken translational symmetry, $n_{-}=2$ and $n_{+}=2$
[Fig.~\ref{fig:fragments}a].  This degeneracy persists for small $U$, but
it is broken as soon as we set a weak imbalance in the hoppings. For any $0
< J_{-}\ll U$, terms proportional to $J_{-}$ connects these states
off-resonantly to excited states like the ones shown in
Fig.~\ref{fig:fragments}c-d. It follows that to second order in
perturbation theory the degeneracy of these states is broken and the energy
of the uniform state, $n_{+}=n_{-}$, is shifted by an amount $\Delta E = -8
(J_-^2/U)\times L$ smaller than the ferromagnetic states with broken
symmetry, $\Delta E=-24 (J_-^2/U) \times L$, which become the true ground
state.

Once we have established that the system spontaneously selects either the
$c_{+}$ or $c_{-}$ order, we can compute the ground state for any value of
$U$ while keeping $J_{-}=0$. Such state will be a product of ground states
of the two-well problem,
\begin{equation}
  \label{frag}
  |\psi\rangle = \prod_{i=1}|\phi^{(\bar n)}\rangle =
  \prod_{i=1}^{L/2} \sum_{m=0}^{\bar n}\phi_{m}^{(\bar n)}
  (c_{2i\sigma}^\dagger)^m
  (c_{2i+1\sigma}^\dagger)^{\bar n-m}|\mathrm{vac}\rangle.
\end{equation}
Here $\bar n = N/L =1$ is the density, $\phi_{m}^{(\bar n)}$ is the
wavefunction of the double-well problem with $\bar n$ atoms and
$\sigma=\pm$ is the selected polarization of the atom.

To delimit the incompressible regions in phase space we only need to
compute the energies of a state with $N=L,L+1$ and $L-1$ particles, which
are variations of the one in Eq.~(\ref{frag}). Since the ground state
energies of a double well with $1$, $2$ and $3$ particles are,
respectively,
\begin{eqnarray}
  \epsilon_1(J_{+},U) &=& -2J_{+},\\
  \epsilon_2(J_{+},U) &=&
  \frac{U}{2}-\sqrt{\left(\frac{U}{2}\right)^2+(2J_{+})^2},
  ~\mathrm{and}\nonumber\\
  \epsilon_3(J_{+},U) &=& 2U-J_{+}-\sqrt{U^2+2UJ_{+}+4J^2_{+}},\nonumber
\end{eqnarray}
the boundaries of the Mott region with $n=1$ particle per site are given by
\begin{eqnarray}
  \mu_p(J'=J) &=& \epsilon_3(2J,U)-\epsilon_2(2J,U),\label{boundaries}\\
  \mu_h(J'=J) &=& \epsilon_2(2J,U)-\epsilon_1(2J,U).\nonumber
\end{eqnarray}
These boundaries are the dash-dot lines we plotted in
Fig.~\ref{fig:dmrg}. A simple inspection of the previous formulas shows
that these curves never cross and become parallel in the limit of $J_{+}/U
\gg 1$.

\begin{figure}[t]
  \includegraphics[width=0.6\linewidth]{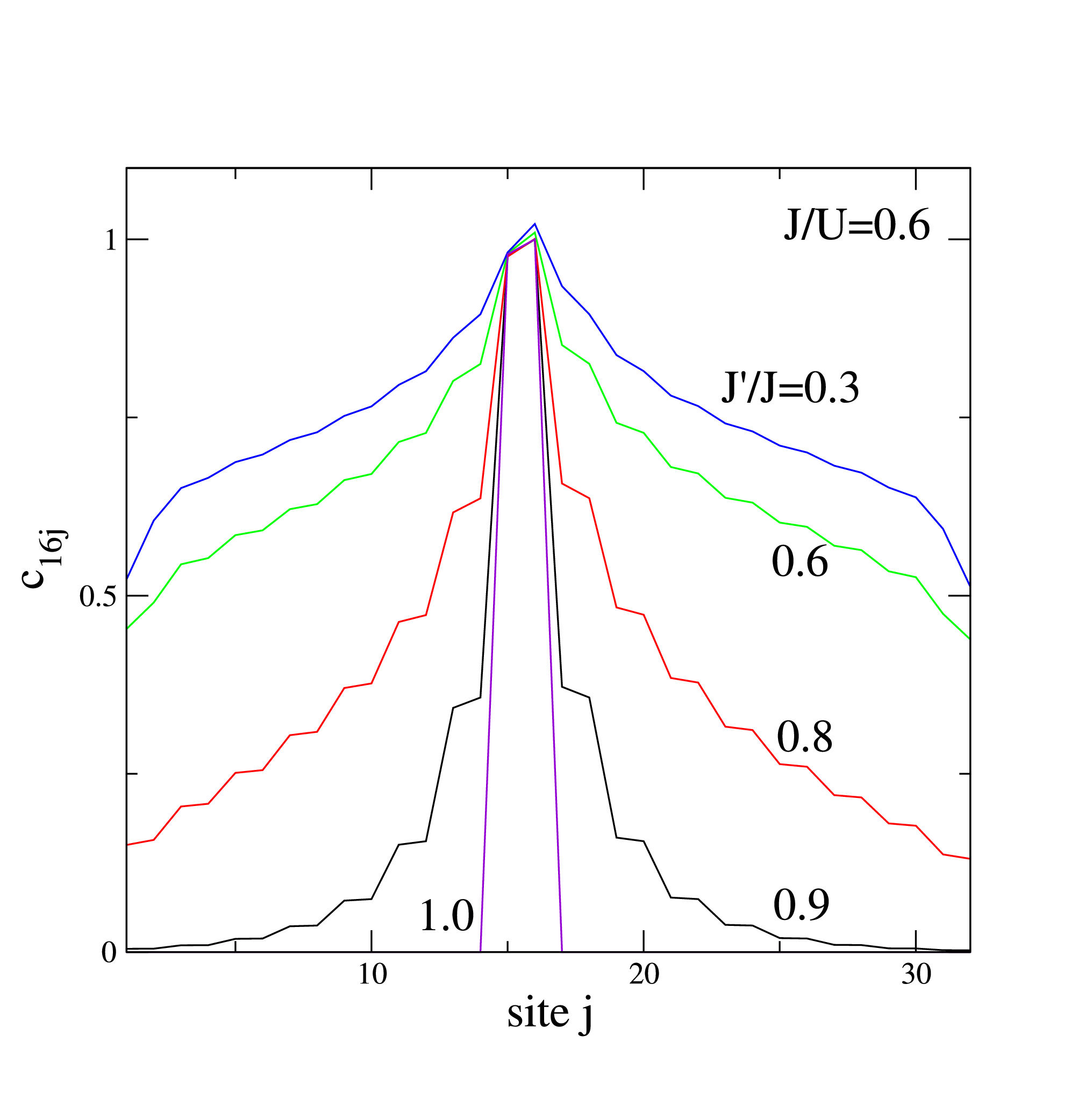}
  \caption{Correlations between different sites, $c_{i,j} := \langle
    c_{i+}^\dagger c_{i+}\rangle$, for $J/U=0.6$ and $L=32$ sites. In solid
    line we plot $c_{16,j}$ for different values of $J'/U$. For the
    strongly coupled sites 15 and 16, the correlations are large. For
    $J\simeq J'$ the correlations are dramatically reduced between sites
    that belong to different double wells in the effective
    superlattice. Note that the correlation function has period two,
    $c_{i,j}=c_{i+2,j+2}$, due to the superlattice structure.}
  \label{fig:site-corr}
\end{figure}

The previous exact results, which are valid for $J_{-}=0$, can be extended
to $J_{-}\ll U,J_{+}$ by means of perturbation theory. In that case there
will be an additional coupling between fragments up to a distance of the
order of the correlation length. The wavefunction can be approximated by
linear combinations of the terms (\ref{frag}) with different occupations of
the double wells. The corrections are of order ${\cal O}(J_{-}/U)$ to the
wavefunction, and of order ${\cal O}(J_{-}^2/U)$ to the energy, but the
dominant term is always the one with equally populated double wells. This
will be further explained in Sect.~\ref{sec:gaps}, where we analyze the
location of the Mott insulator to superfluid transition. That this picture
is applicable to a great degree is appreciated in Fig.~\ref{fig:site-corr},
where we plot the correlations between different sites. For $J'=0.9$, sites
connected by $J_{+}$ are strongly correlated, while the correlations
decrease significantly between different pairs of sites.

\subsection{Hard-core bosons}
\label{sec:hard-core}

To understand the behavior of the bosons in the strongly interacting limit,
$U\gg J_{\pm}$ it is useful first to see what happens if there are no
frustrating terms, $J'=0$.  We know that on the one hand, if we are below
half-filling, $N <L$, the gas is a Tomonaga-Luttinger liquid with
spin-charge separation \cite{voit95}. In the limit of $U\to\infty$ the spin
degrees of freedom become degenerate and the whole system behaves like free
fermions or a Tonks gas. On the other hand, for half filling, $N=L$, the
sample becomes a Mott for any strong interaction $U\gg J$. The atoms cannot
move but their spin degrees of freedom interact with a ferromagnetic
Heisenberg interaction of strength $t=-J^2/U$ and a gapless spectrum made
of spin waves.

The diagonal coupling $J'$ changes this landscape slightly. For half
filling, $N=L$, we still have a Mott phase, with one atom per
site. Connecting with the previous section, as we increase $U$ the
wavefunction of the two-well fragment (\ref{frag}) becomes closer and
closer to $\phi_{m}^{(\bar n)}=\delta_{m,\bar{n}/2}$. This is a smooth
evolution which is why we speak of a crossover. Furthermore, compared to
the unfrustrated case $J'=0$, the spin rotational symmetry is now broken,
favoring the $X$ direction or $c_{\pm}$ states in the second quantization
language. The effective spin model is, up to constants, a ferromagnetic one
\begin{equation}
  \label{XXZ}
  H_{\mathrm{spin}} =
  -\frac{1}{U} \sum_k \left[ 2J'^2\sigma^x_k\sigma^x_{k+1}
    + J_+J_-\vec{\sigma}_k\vec{\sigma}_{k+1}\right],
\end{equation}
and spin excitations are prevented by an energy gap $\Delta = 2J'^2/U$
\cite{koma97}. This gap is evident also in Fig.~\ref{fig:gap}a, in the
region of small hopping, $J\ll U$.

\begin{figure}[t]
  \includegraphics[width=0.6\linewidth]{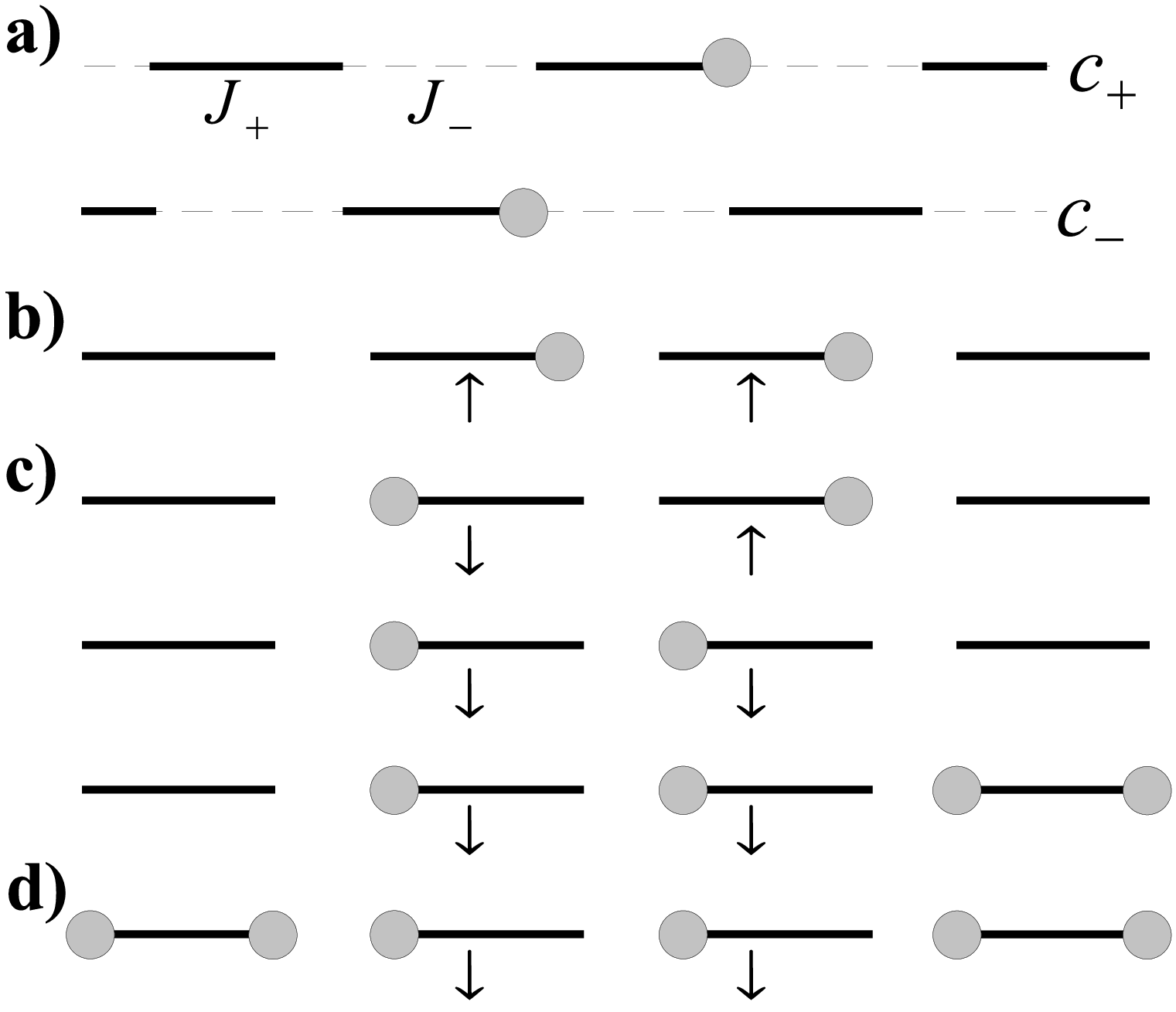}
  \caption{(a) A possible state of the ladder, where the circles
    denote the location of a particle. (b) The same state mapped to a
    1D problem for $J_-=0$. Allowed (b)-(c) and forbidden (d) states
    in the hard-core bosons limit ($U\to\infty$).}
  \label{fig:mott}
\end{figure}

For smaller fillings, i.~e. $N<L$, the previous spin model is no longer
valid because the chain has holes. Nevertheless one can easily compute
things if $J_-=0$. The ground state manifold is composed of variations of
the fragmented wavefunction (\ref{frag}). In particular, the set of
available configurations is shown in Fig.~\ref{fig:mott}, where we show
three kinds of bonds. Those with zero or two particle, which are frozen,
and bonds with one particle on the left or right sites, which can be
identified with spins. These effective spins form ferromagnetic blocks
which are separated by the frozen sites. If the block is surrounded by at
most one doubly occupied bond, then it can only adopt one state and it has
zero energy [Fig.~\ref{fig:mott}d]. If the block, on the other hand is
surrounded by empty blocks it may host at most one domain wall
[Fig.~\ref{fig:mott}c] with some small negative energy due to the motion of
the domain wall (for instance, the states in Fig.~\ref{fig:mott}b and
Fig.~\ref{fig:mott}c are connected by hopping). The effective Hamiltonian
for a block of size $M$ is
\begin{equation}
  H_{DW} = -J \sum_{m=0}^M (|m+1\rangle\langle m|+|m\rangle\langle m+1|),
\end{equation}
where $m$ denotes the number of particles to the left. This gives an
energy spectrum $\epsilon_k(M) = -2J\cos[k\pi/(M+2)]$, for $k=1\ldots
M+1$. Since $\epsilon_k(M)$ is a decreasing, convex function of $M$,
the lowest energy state is achieved by making $L-N+1$ blocks with
about $N/(L-N+1)$ atoms each. The excitations are still gapped with
$\Delta \propto J$, and they can be of two types: excitations of the
domain wall, and merging of two blocks. If $N < L/2$ the latter are
more important and of order $\Delta = 2 e_1(1) - e_1(2) \simeq 0.5858
J$.  On the other hand, if $N>L/2$ the excitation energy of a domain
wall can be pretty small, ${\cal O}(1/M^2)$.

\subsection{Scaling of the Mott transition}
\label{sec:gaps}

\begin{figure}
  \includegraphics[width=0.5\linewidth]{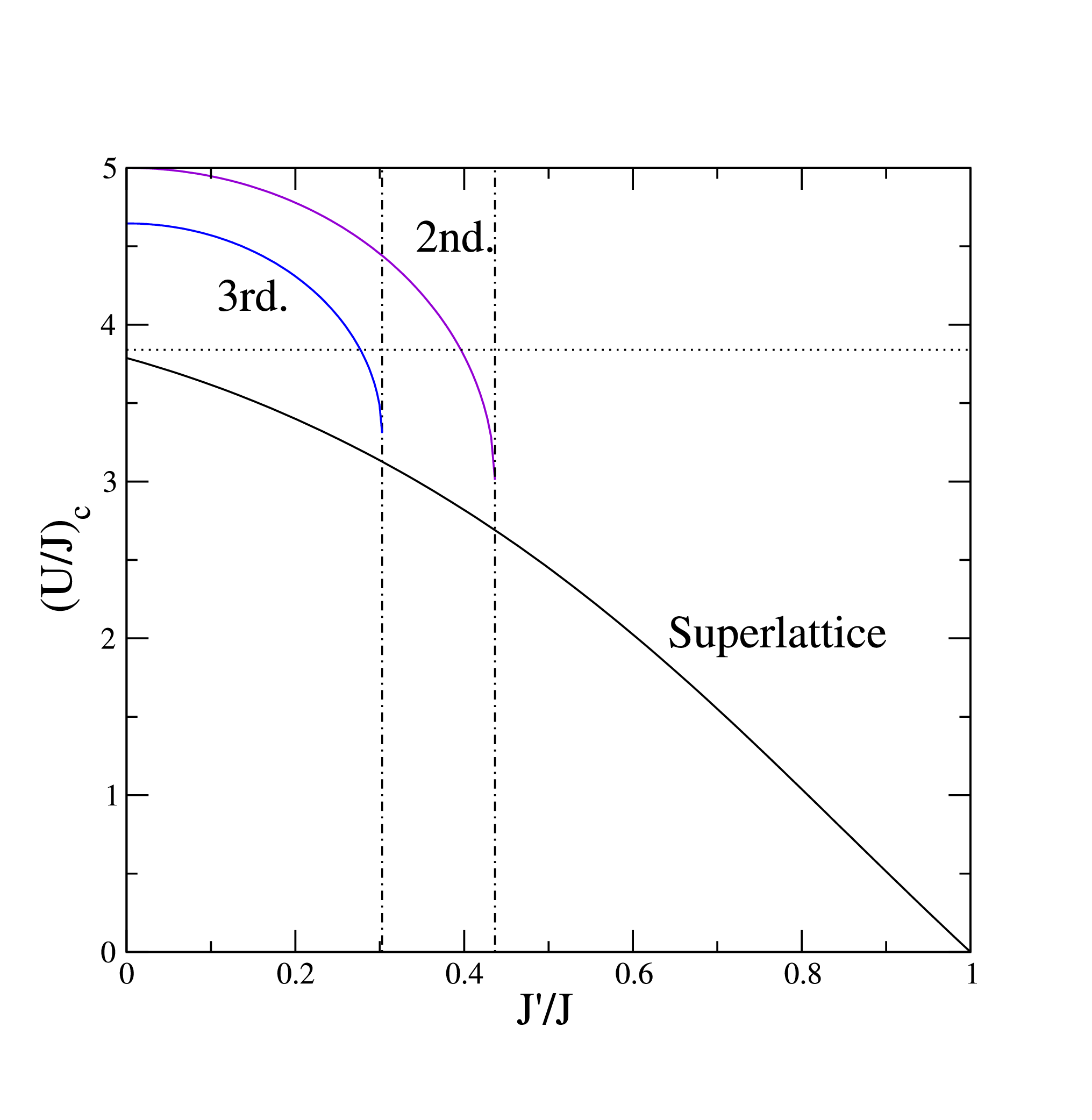}%

  \caption{Critical value of $U/J$ at which the Mott insulator to
    superfluid transition should take place. The blue and violet curves are
    based on the strong coupling expansion of third and second order in
    $(J/U)$ developed in Ref.~\cite{freericks96}. The solid line is based
    on a different strong coupling expansion of order ${\cal
      O}(|J-J'|^2/U)$, developed below. The dotted line is the critical
    value $U_c=3.84J$ computed with DMRG for unfrustrated models
    \cite{kuehner98}. Note the agreement between the two latter methods for
    $J'=0$.}
\label{fig:critical}
\end{figure}

In this subsection we attempt to analyze how the superfluid to insulator
transition changes due to frustration. We will do it in two different
perturbative methods. The first one is based on a strong coupling expansion
around $J,J'=0$, while the second one is based on a strong coupling
expansion around $J'=J$. Remarkably, this second method, which is more
accurate than the trivial strong coupling expansion \cite{freericks96}, may
be contemplated as a real space renormalization technique in which the
alternating Hubbard model is replaced by an effective one with a
renormalized hopping and half the sites.

Our first attempt starts with the estimates of the critical value of $J/U$
given in Ref.~\cite{freericks96}. Applying perturbation theory up to ${\cal
  O}(J^3/U^3)$, it is possible estimate the change in the energy induced by
a small amount of hopping. The resulting energies can then be used to
compute the particle and hole excitation gaps, and delimit the borders of
the incompressible regions in the hopping vs. chemical potential parameter
space, as in Fig.~\ref{fig:dmrg}. To apply the theory in
Ref.~\cite{freericks96} we need the hopping matrix\footnote{We assume
  periodic boundary conditions.}
\begin{equation}
  -t_{ij} = -J_i \delta_{i+1,j} - J_{i-1} \delta_{i-1,j},
\end{equation}
the smallest eigenvalue of $(-t_{ij})$
\begin{equation}
  \lambda_{\mathrm{min}} = - \max\{|J_{+}-J_{-}|,|J_{+}+J_{-}|\} = -2\max\{J,J'\},
\end{equation}
the wavefunction of the single-particle ground state, $f_i = 1/\sqrt{L}$,
and finally the sums
\begin{equation}
  \sum_{ij} t_{ij}^2 f_j^2 = J_{+}^2 + J_{-}^2,\quad
  \sum_{ij} f_i t_{ij}^3 f_j = J_{+}^3 + J_{-}^3.
\end{equation}
As shown in Fig.~\ref{fig:critical}, the transition does indeed shift to
larger values of $J/U$ as we switch on the diagonal coupling $J'$. However,
the prediction is not very consistent between second and third order
expansions, and it even fails to provide the right value for $J'=0$,
probably due to the asymptotic nature of these expansions.

Our second attempt focuses on the regime of strong frustration or values of
$J'$ close to $J$. We have seen that for $J'=J$ the incompressible regions
fill the entire phase space. With very simple arguments it is possible also
to compute how they shrink for nonzero $|J'-J|=\epsilon\ll U$. The
reasoning begins by noticing that our system is a superlattice, where the
weak hopping terms can be treated perturbatively. However, instead of using
the Mott states as unperturbed states one must use the exact ones for
$J_{+} > 0$. Basically, we have an unperturbed state with energy
$E=L\epsilon_2/2$ and several unperturbed manifolds above it, consisting on
holes and excitations in the superlattice, with energy gaps of order
$\epsilon_3-\epsilon_2$, $\epsilon_1-\epsilon_2$. We can compute the
corrected $\mu_p$ and $\mu_h$ as a function of $J_{-}$. The equation for
the critical point then looks as follows
\begin{equation}
  \label{criticalU}
  \mu_p - \mu_h = \epsilon_3(U)+\epsilon_1(U)-2\epsilon_2(U)
  - J_{-} c(U) + {\cal O}(J_{-}^2) = 0,
\end{equation}
where the energies and the matrix elements of the hopping Hamiltonian,
$c(U)$, are functions of $U$. We have solved numerically the highly
nonlinear equation (\ref{criticalU}) for $U$ as a function of $J_{-}$ and
$J_{+}$. The results are plotted in Fig.~\ref{fig:critical}. As seen there,
this second method interpolates properly between the critical value of
$U=3.84$ obtained by DMRG for the unfrustrated model \cite{kuehner98}, and
the critical value $U=0$ computed for $J=J'$. Furthermore, the whole line
deviates very little from $U_c = U_c(J'=0)|J-J'|$.

\subsection{Correlators}
\label{sec:correlators}

As it will be useful to identify the different phases experimentally, we
have computed the single-body and two-body correlation functions in the
three regimes. The most interesting values are the single-body correlators,
$\langle a_i^\dagger a_j\rangle$. In the superfluid limit, due to the
one-dimensional character of our problem, we expect that correlations decay
algebraically. In the fragmented regime, we have two possible ground
states, depending on whether the fragments sit on even or odd junctions,
which is equivalent to specify whether the fragments are made of $c_+$ or
$c_-$ particles. In the first case we will have
\begin{equation}
  \langle c_{2i,+}^\dagger c_{2i,+}\rangle =
  \langle c_{2i,+}^\dagger c_{2i+1,+}\rangle =
  \langle c_{2i,+}^\dagger c_{2i,+}\rangle = 1/2,
\end{equation}
with all other correlators being zero, and in the $c_-$ case we will have
the same correlation pattern but displaced by one lattice site. Finally, in
the Mott insulator regime we have no correlations and $\langle
c_{i\pm}^\dagger c_{j\pm}\rangle_{\mathrm{frag}} \sim \delta_{i-j}.$

In experiments what is measured is the time of flight images. These
pictures are related to the momentum distribution of the sample or the
Fourier transform of the single-particle correlations
\begin{equation}
  n_q \propto \sum_{k,l=1}^{L} e^{i2\pi q(k-j)/L}
  \langle a_k^\dagger a_j\rangle\,\quad q \in [-L/2,L/2].
\end{equation}
This function is peaked at $q=0$ and decays towards the sides of the
Brillouin zone. The half-width of the peak is related to the correlation
length of the sample. In the superfluid region it will be proportional to
the size of the lattice and limited by finite temperature effects
\cite{cazalilla06}. As soon as we cross the phase transition towards the
fragmented phase, though, the correlation length decays abruptly to $\xi =
2$ sites and the previous function has a constant width
\begin{equation}
  n_q \sim 1 + \cos(2\pi q/L)
\end{equation}
From there on we expect a smooth crossover towards $n_q=1$ which is the
Mott-Insulator regime. All this phenomenology is evident in
Fig.~\ref{fig:interference}.2

\section{Villain model}
\label{sec:villain}

The other limiting case in our setup is
that of vertical frustrating interactions, a configuration resembling
the \textit{odd model} or Villain lattice \cite{diep04}. Once more it
is convenient to change basis as in Eq.~(\ref{rotation}), so that the
frustrating terms become alternating energy shifts
\begin{equation}
  H = \sum_{i;\sigma=\pm} \left[
  -J (c^\dagger_{i\sigma} c_{i+1\sigma} + \mathrm{H.c.})
  - (-1)^i \sigma J' c^\dagger_{i\sigma}c_{i\sigma}
  + \frac{U}{2} n_{i}^2\right].
  \label{modeld}
\end{equation}
The effect of the Raman coupling is now equivalent to a superlattice.  This
additional potential splits the Wannier band into two effective bands
separated by a gap $J'$, each band having a reduced width $\Delta J
:=(\sqrt{4J^2+J'^2}-J')/2$. The superlattice localizes the atoms on
alternating rungs, forming an antiferromagnetic configuration.  Instead of
exhibiting fragmentation, the system goes straight from a superfluid to a
Mott-insulator, but now the transition happens for weaker interactions,
$U\sim \Delta J$.

\section{Experimental realization}
\label{sec:experiments}

There are several issues that one has to consider when implementing our
Hubbard models using neutral atoms. The first one is how to relate $U$, $J$
and $J'$ with experimental parameters such as the intensity of the Raman
laser and the strength of the optical lattice.  Additionally one has to
consider how to prepare the ground state and how cold the sample should
be. Finally one has to think about procedures to detect the different
phases. We will address all of these questions in the following sections.

\subsection{Microscopic theory of Hubbard model}

\label{sec:wannier}

In this subsection we show how to derive the constants $U$, $J$ and
$J'$ using a band structure calculation that takes into account the
Raman coupling, $\Omega(x) = \Omega_0\cos(kx+\phi)$, and the confining
lattice,
\begin{equation}
  V(x)= V_{0x} \sin(kx)^2+ V_{0\perp} [\sin(ky)^2 + \sin(kz)^2].
\end{equation}
Our work generalizes the ideas from Ref.~\cite{jaksch98}, but now our
microscopic Hamiltonian
\begin{equation}
  H = \sum_{\sigma,\sigma'}
  \int d\mathbf{x}\, \psi_\sigma^\dagger(\mathbf{x}) \left[
    -\frac{\hbar^2}{2m}\nabla^2 +
    W_{\sigma\sigma'}(\mathbf{x})\right]\psi_{\sigma'}(\mathbf{x})
  \label{fullH}
\end{equation}
contains a potential term
\begin{equation}
  W_{\sigma\sigma'} = V(\mathbf{x}) \delta_{\sigma\sigma'}
    + \Omega(\mathbf{x}) \delta_{\sigma\bar{\sigma'}}
    + \sum_{\alpha=\uparrow,\downarrow} \frac{\tilde{U}_{\sigma\alpha}}{2}
    |\psi_\alpha(\mathbf{x})|^2
\end{equation}
that includes not only the trapping potential, but the coupling and
the interaction between atomic states.

Following \cite{jaksch98}, we perform a tight-binding approximation
and replace the bosonic operator by an expansion of the form
\begin{equation}
  \psi(\mathbf{x}) \simeq \sum_{i,j,k,\sigma}
  a_{ijk\sigma} w_x(x-b i)w_y(y-bj)w_z(z-bk),
\end{equation}
where $b=\pi/k$ is the period of the lattice and $w_{x,y,z}$ are the
Wannier function corresponding to the lowest Bloch band of the
lattice, along the X, Y and Z directions, respectively. By
substituting this expansion into the full Hamiltonian (\ref{fullH})
and retaining the most important terms, one arrives to a Bose-Hubbard
model (\ref{Hamiltonian}).

To compute the parameters of our model, we will assume that the confinement
along the transverse directions, $V_{0\perp}$, is big enough, so that
hopping is negligible and our system is made of multiple, disconnected 1D
lattices. However, even with this assumption there are still too many
``knobs'' to tune. To simplify the problem it is customary to
adimensionalize all magnitudes using the energy of the lattice, $E_R=\hbar^2
k^2/2m$, and its period, $b$, as units. We replace the trapping potential
with $V \sim V_0 \cos(2kx)$, make the change of variables $2kx \to 2\pi x$
and rescale the wavefunction accordingly, $w\to w\sqrt{k/\pi}$. As a result
we obtain formulas for the hopping
\begin{equation}
 \frac{J}{E_R}=
 \int dx\,w_x(x+1)\left[-\frac{1}{\pi^2}\frac{d^2}{dx^2} + \frac{V_0}
 {2E_R}\cos(x)\right]w_x(x),
\end{equation}
the Raman coupling
\begin{equation}
  \frac{J'}{E_R} = \int \frac{\Omega_0}{E_R}\cos(\pi x+\phi)|w_x(x)|^2
  dx,\quad \phi\in\{0,\pi/2\}
\end{equation}
and the interaction energy
\begin{equation}
  \frac{U_{\sigma\sigma'}}{E_R} =
  a_{\sigma\sigma'}^{(1D)} \int dx |w_x(x)|^4 =:
  a_{\sigma\sigma'}^{(1D)} I(V_0/E_R).
\end{equation}
Notice that $w_x(x)$ is now a Wannier function computed for a single,
one-dimensional problem with period $2\pi$ and trapping $V_0/E_R$. The only
other free parameters are $\Omega_0/E_R$, and the effective one-dimensional
scattering length~\footnote{Note that this is an
  effective value, unrelated to the scattering theory of cold atoms in
  1D tubes by Olshanii \cite{olshanii98}.}
\begin{equation}
  a_{\sigma\sigma'}^{(1D)} := \frac{4}{\pi^2} k a_{\sigma\sigma'}^{(3D)}
 I\left(V_{0\perp} / E_R \right)^2.
\end{equation}
At this point one has to solve a family of one-dimensional problem with the
periodic potential $\cos(2\pi x)$ in order to obtain the relative values of
$J, J'$ and $U$.

\begin{figure}[t]
  \centering
  \includegraphics[width=0.6\linewidth]{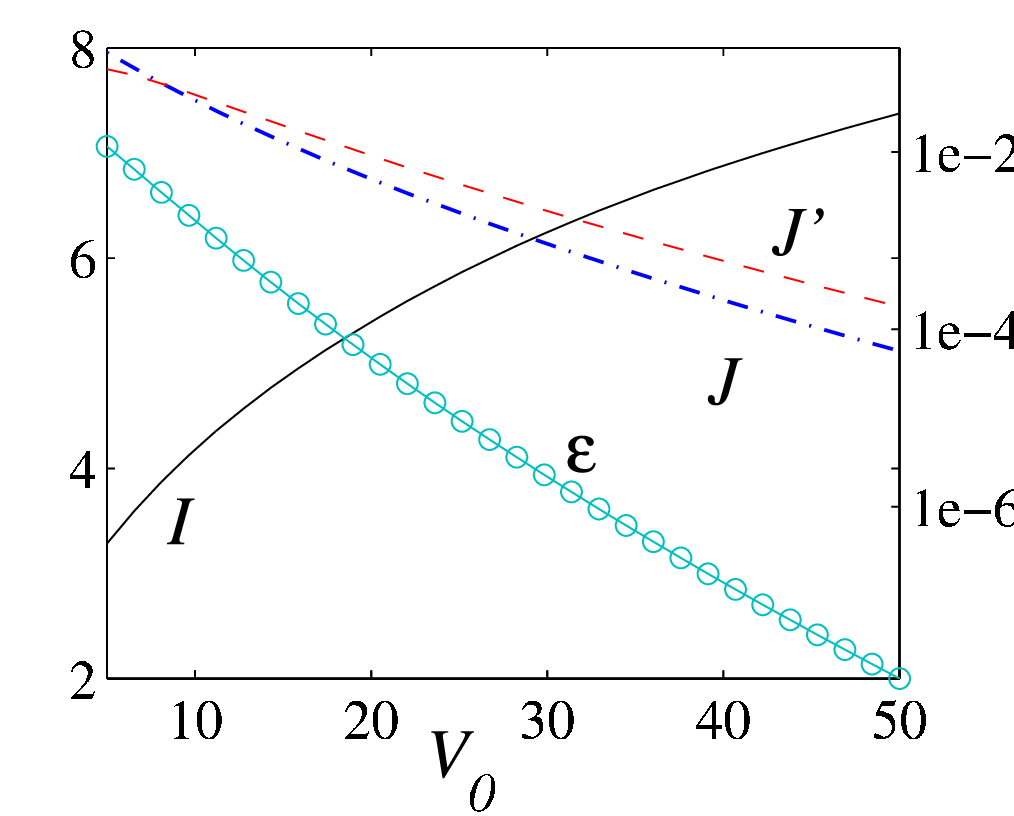}
  \caption{Parameters of the model, $I$ (solid, left axis), $J$
    (dash-dot, right) and $J'$ (dashed, right) as a function of the
    lattice depth, $V_0$, for a Rabi coupling $\max|\Omega(x)|=V_0/2$,
    and diagonal interactions, $\phi=\pi/2$.  The value $\varepsilon$
    (circles, right) is the effective coupling to higher bands
    (\ref{coupling}). All parameters are measured in units of the recoil
    energy of the confining lattice.}
  \label{fig:wannier}
\end{figure}

\subsection{Experimental parameters}

\label{sec:parameters}

From the computations sketched in Sect.~\ref{sec:wannier}, it follows that
if we choose the Rabi coupling of similar strength to the confinement,
$\Omega_0 = V_0/2$, then the ordinary hopping, $J$, and the ladder coupling,
$J'$, are also of the same order of magnitude [Fig.~\ref{fig:wannier}].
That means that by slightly tuning $\Omega_0$ it should be experimentally
feasible to cover all possible regimes, from $J_-=J_+$ to $J_-=0$.

Now it only remains the question of whether we can make $U/J$ small or big
enough to explore the different phases. For a typical experiment with
$^{87}$Rb, we have $a^{3D}\sim5$nm and the wavelength of the laser is
$\lambda\sim$850nm, so that $a_{\sigma\sigma'}^{(1D)} \sim 1.8\times 10^{-3}
I^2$. Now, for a strong confinement we have that $I$ ranges from 2 to 8, and
comparing with the evolution of $J$ and $J'$ in Fig.~\ref{fig:wannier} we
see that the value of the interaction can indeed change from a superfluid
regime, $U < |J-J'|$, to the Mott insulator regime, $|J+J'|\ll U$. In other
words, if we keep $J-J'$ not too small, the laser intensities will not
differ much from those used in current experiments
\cite{paredes04,stoferle04} and we will be able to generate all phases.

While the validity of the tight-binding approximation is well established
for the ordinary Bose-Hubbard model, we have introduced a coupling between
neighboring sites that might excite the atoms to higher Bloch bands. We have
studied this effect and computed the following effective strength of the
coupling
\begin{equation}
  \label{coupling}
  \varepsilon := \left|\frac{\max |Q_i(\Omega_0/E_R)|}{\Delta E}\right|^2,
\end{equation}
which is a function of the energy difference between bands, $\Delta E$, and the
coupling between Wannier functions of the lowest and first excited
bands
\begin{equation}
  Q_i(\Omega_0/E_R) := \frac{\Omega_0}{E_R}
  \int w_x(x) \cos(x/2+\phi) w_x^{(2)}(x-i).
\end{equation}
When small, the value $\varepsilon$ measures how much mixing of the excited
wavefunctions there is in the ground state and it is a function of
$\Omega_0$. As Fig.~\ref{fig:wannier} shows, for the diagonal interactions
($\phi=\pi/2$) the coupling is indeed extremely small, $\varepsilon <
10^{-2}$, and we can be sure that our approximations are valid.  Similar
results are obtained for $\phi=0$.

\subsection{Preparation and observation}
\label{sec:preparation}

\begin{figure}[t]
  \centering 
  \includegraphics[width=0.6\linewidth]{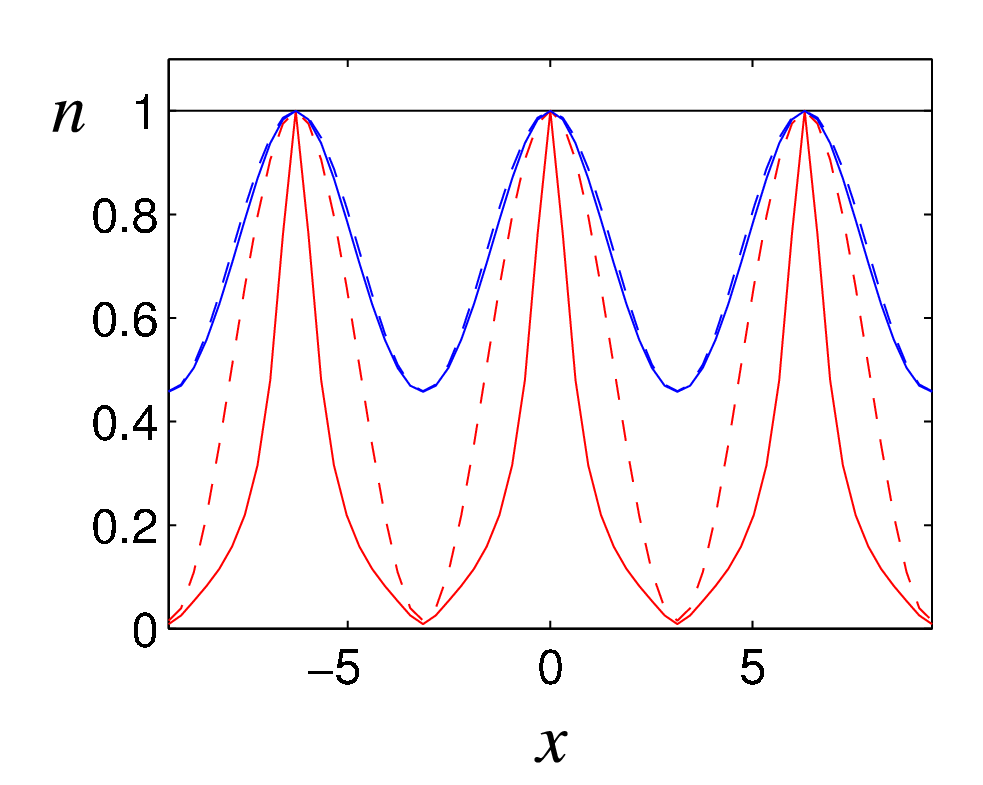}
  \caption{Ideal interference pattern arising from time of flight
    images of the model with diagonal interactions (\ref{diagonal}). We
    show results for $N=L=32$ particles, $J/U=0,0.05$ and $0.5$, from
    top to bottom, using $J'=J$ (dash) and $J'=0.9J$ (solid). See
    Sect.~\ref{sec:preparation} for details about $n$ and $x$.}
  \label{fig:interference}
\end{figure}

The remaining issues are related to the possibility of preparing and
unambiguously identifying the ground states of our model
(\ref{Hamiltonian}). The state preparation can be done in a two step
process. One begins with a 1D superfluid and all atoms in the same internal
state, and slowly increasing the lattice depth. Judging from current
experiments, this should be a robust way to create a Mott or Tonks gas
phase \cite{paredes04}. For $U\gg J,J'$ the density excitations will be
frozen and the dynamics will be ruled by the effective ferromagnetic model
(\ref{XXZ}). One can can then slowly increase $J'$ to create the ground
state of (\ref{XXZ}) which will be the ground state of the frustrated
model.

Regarding detection, the three phases of model (\ref{diagonal}) can be
identified from time of flight images. First of all, for $U \ll
J_{\pm}$ we expect a quasi-condensate which will produce strong
interference peaks in either of the atomic species. The width of these
peaks may be slightly modified by temperature \cite{cazalilla06} but
it should remain approximately constant for $U<J_{\pm}$. For
moderately strong interactions, $J_{-} < U < J_{+}$, the correlation
length will decrease substantially as we expect coherences only inside
each fragment. Thus according to Sect.~\ref{sec:correlators} the
interference pattern will be approximately
\begin{equation}
  |\psi(tp/m)|^2 \sim |\tilde{w}(p)|^2 [1 + \cos(p)] =:
  |\tilde{w}(p)|^2 n(p).
\end{equation}
Here $\tilde w(p)$ is the Fourier transform of a Wannier wavefunction, $n$
is the ideal interference pattern and the position $x=tp/m$ is related to
the time, $t$, the mass of the atoms, $m$, and its momentum in the lattice,
$p$. In practice, for $J'\neq J$, the remaining coherence between fragments
can modify this pattern, giving thinner peaks, as shown in
Fig.~\ref{fig:interference}.

Another signature of the transition from quasi-condensate to fragmented
phase is a spontaneous symmetry break that makes the ground state be formed
of either $c_+$ or $c_-$ particles. The energy gap in this phase is related
to the stiffness of the atoms with respect to a change in the internal
state and it could be probed spectroscopically.

A third signature which applies to both the fragmented phases and the
ground state of the Villain model is the existence of magnetic order.
More precisely, for strong interactions $J_+ < U$, the ground states
of the diagonal and Villain frustrated models have ferro- and
antiferromagnetic order, respectively. This pattern reveals itself in
high order correlations between the number of atoms of $c_{+}$ and
$c_{-}$ at different sites: $ \langle
c^\dagger_{j\sigma'}c_{j\sigma'}c^\dagger_{i\sigma}c_{i\sigma}\rangle
\propto1+ \epsilon \sigma \sigma' (-1)^{i-j}$, where $\epsilon=1$ or
$0$ for each model, respectively.  To observe these correlations one
must first rotate the atoms with a $\pi/2$ pulse, as in
Eq.~(\ref{rotation}), and afterwards analyze the quantum fluctuations
in the time of flight images \cite{foelling05}. Alternatively, if the
experiment allows for state dependent lattices and photo-association,
it will be advantageous to use the techniques suggested in
\cite{garciaripoll04} to measure the number correlations. This second
method should indeed provide a much stronger signal.

In current experiments we have to consider two additional sources of
imperfection that can influence our observations. One of them is the
residual harmonic confinement which is always present due to the
difference on intensity along the lasers that trap the atoms. From the
diagram of Fig.~\ref{fig:dmrg}, the harmonic confinement will cause
the atomic cloud to be made of nested insulating shells with different
densities \cite{jaksch98}, while the size of the superfluid regions
will substantially decrease and strictly vanish for $J=J'$. This
structure can cause noise in the interference patterns, but, as it has
been shown in recent experiments \cite{folling06}, it can also be
probed using RF knifes and spin-changing collisions.

The second source of imperfections is temperature, which will influence the
measurements in two ways. First of all, for $T\neq 0$, the weakly
interacting region will not be a condensate, but a quasi-condensate, where
phase fluctuations are energetically cheap but the system remains
superfluid. One might argue that these phase fluctuations will destroy the
interference peaks, but as we have seen in 1D experiments with optical
lattices \cite{stoferle04}, this does not seem to happen in practice.
Moreover, the correlation length of the quasi-condensate will be much larger
than that of the fragmented phase, a fact that will be evident in the
interference pattern. It remains the question of whether the fragmented
phase itself is robust against temperature and we argue that indeed it is.
This phase will be observed if the temperature of the system is smaller or
comparable to the energy gap, $\Delta$. But as shown in Fig.~\ref{fig:gap}
the energy gap can be considerably large, $\Delta \sim 0.6 J$, sufficiently
larger than the temperature limitations of the latest experiments
\cite{paredes04,folling06}. We expect that under these conditions the
fragmented phase can be unambiguously identified.

\section{Conclusions}

\label{sec:conclusions}

In this work we have introduced several new ideas. The first one is to
use internal states of atoms in an optical lattice to simulate
additional spatial dimension and to implement ladders. While it may be
argued that there are previous works which have actively employed the
internal degrees of freedom of the atom for quantum simulation
\cite{jaksch03,osterloh05,jane03,duan03,yip03,garciaripoll04}, it is
in this work that they are used to effectively increase the
dimensionality of the lattice.

The second idea is to couple the atomic degrees of freedom in a spatially
dependent way \cite{jaksch03}, so as to induce a frustrating hopping in the
new virtual dimensions. The frustrating nature if this assisted tunneling
arises from the sign of the Raman coupling and it is best appreciated in the
hard-core bosons limit, in which the model is tantamount to a spin
Hamiltonian [Sect.~\ref{sec:frustration}]. Additionally, this frustrating
hopping can also be seen as originating multiple small Mach-Zender
interferometers in the lattice, such that destructive interference disturb
the motion of atoms along the lattice [Fig.~\ref{fig:fragments}e].

The most dramatic consequence of these two kinds of tunneling is the
breakdown of superfluidity. The ground state fragments into a macroscopic
number of double-well subsystems which lose coherence among them as we
increase the strength of interactions. This situation is reminiscent of a
Bose glass where a disordered energy landscape creates islands of
superfluid regions with no coherence among them~\cite{fisher89}.  However,
in our case the fragmentation is induced by frustration and not by
disorder.

It is possible to qualitatively connect the Physics observed here with that
of other spin models. On the one hand, the modulation of the hopping leads
to an effective supperlatice in the internal plus motional degrees of
freedom. This superlattice gives rise to a larger coherence between
neighboring lattice sites, similar to what happens in the spin-Peierls
effect, where a modulation of a Heisenberg interaction leads to
dimerization \cite{giamarchi04}. Thus, in contrast to other models where
dimerization happens spontaneously \cite{haldane82}, here it is imposted by
the quantum interference of hopping terms.

However, while this analogy to spin models explains the fragmentation, our
system goes beyond this simple picture, since we actually have two
superlattices which are connected by repulsive interactions. The internal
degrees of freedom of the atoms give rise to rich Physics both in the
strongly repulsive regime and in the insulator to superfluid transition. In
particular, the dynamics of the spin deep in the superfluid region deserve
tocc be studied in future work.

The methods shown here can be applied to simulating spin ladders as the
ones studied in Ref.~\cite{garciaripoll04}. Additionally, the internal
degrees of freedom of the atoms can be used to change the topology of the
lattice from square to triangular. Finally, a natural extension would be to
combine the Raman coupling used here with a two-dimensional or even a
three-dimensional optical lattice, so as to simulate new topologies and
other frustrated bilayer magnetic models that may have not been considered
so far.

\section*{References}

\bibliographystyle{unsrt}
\bibliography{atoms}

\end{document}